\documentstyle[12pt]{article}
\newcommand{\rf}[1]{(\ref{#1})}
\newcommand{\beq}{\begin{equation}}
\newcommand{\eeq}{\end{equation}}
\newcommand{\bea}{\begin{eqnarray}}
\newcommand{\eea}{\end{eqnarray}}
\newcommand{\g}{\gamma}

\newcommand{\m}{\mu}

\newcommand{\Del}{\Delta}

\newcommand{\oh}{\frac{1}{2}}
\newcommand{\oq}{\frac{1}{4}}

\newcommand{\ra}{\rangle}
\newcommand{\la}{\langle}

\begin{document}
\topmargin 0pt
\oddsidemargin 5mm
\headheight 0pt
\headsep 0pt
\topskip 9mm

\hfill    NBI-HE-99-36

\hfill September 1999

\begin{center}
\vspace{24pt}
{\large \bf Branched Polymers Re-Revisited}

\vspace{24pt}

{\sl J. Ambj\o rn }$^a$, {\sl B. Durhuus }$^b$ and {\sl T. Jonsson }$^c$
\vspace{24pt}

$^a$~The Niels Bohr Institute\\
Blegdamsvej 17, DK-2100 Copenhagen \O , Denmark\\         
{\it email ambjorn@nbi.dk}
\vspace{24pt}

$^b$~Matematical Institute,\\
University of Copenhagen\\
Universitetsparken 5, DK-2100 Copenhagen \O, Denmark\\
{\it email durhuus@math.ku.dk}
\vspace{24pt}

$^c$~Science Institute,\\
University of Iceland,\\
Dunhagi 3, IS-107 Reykjavik, Iceland,\\
{\it email thjons@raunvis.hi.is}

\end{center}

\vfill

\begin{center}
{\bf Abstract}
\end{center}

\vspace{12pt}

\noindent

We point out some misconceptions in a recent paper by H. Aoki et al.  In
particular, the claim that the two-point function of branched polymers
behaves as $p^{-4}$ instead of $p^{-2}$ for large $p$ is
mistaken and in no way a precondition for the Hausdorff dimension of
branched polymers having the well known value four.

\vfill

\newpage

\section{Introduction}

Branched polymers provide us with one of the simplest 
generalizations of random walk. The theory of branched 
polymers is well known (see e.g.\ \cite{book} and references 
therein).

In a recent paper \cite{kawai} it is claimed that the standard treatment of
the theory of branched polymers, as 
presented for instance in \cite{book}, is not correct.
In this note we show that this claim is false.

The problem under discussion concerns the propagation of branched polymers 
in target space.  This is analogous to the propagation 
of a particle in space-time or the propagation of a bosonic string in 
26-dimensional space-time. The grand canonical ensemble provides a convenient
framework for developing the theory and this is how it is done 
in \cite{book} and in
many of the papers referred to therein. This approach is 
claimed to be erroneous in \cite{kawai}, where the discussion is based on the
canonical ensemble.  As we explain below, the basic
requirement of the authors of \cite{kawai} 
on the relation between the two approaches, namely that  
{\it the correlation 
functions in the grand canonical ensemble at the critical value of fugacity
should reproduce those in the canonical ensemble for large N:}
\beq\label{1}
\lim_{N\to \infty} \frac{G_N}{Z_N} = \lim_{\m\to \m_c} 
\frac{G_\m}{Z_\m},
\eeq
is not relevant, and in fact incorrect.

In addition, we explain that the three main conclusions of \cite{kawai} 
i.e.\ (i) that the 2-point function 
$G_\m(p)$ differentiated w.r.t.\ the ``chemical''  potential $\m$ 
behaves like $1/p^4$ for large $p$, (ii) that the 3-point function 
behaves like 
\beq\label{2}
G^3_\m(p,q) \sim G_\m(p)G_\m(q)G_\m(p+q),
\eeq
and (iii) that the 3-point function differentiated w.r.t.\ $\m$ is given by
\beq\label{3}
\frac{d G^3_\m(p,q)}{d\m} \sim G'_\m(p)G_\m(q)G_\m(p+q)+
G_\m(p)G'_\m(q)G_\m(p+q) + G_\m(p)G_\m(q)G'_\m(p+q),
\eeq
(Eqs.\ (47-48) in \cite{kawai}), are all obviously correct, and in fact trivial
consequences of what the authors call the ``naive'' approach. 
However, their claim that only the differentiated three point function 
is consistent with a
correct ``thermodynamical'' limit is misleading. Eq. \rf{2} {\it is} the
correct  3-point function, defined in the same way as one would define 
a 3-point function in string theory, and it does not contain any non-universal
part. Eq. \rf{3} is {\it not} the 3-point function but (by definition)
the derivative of the 3-point function with respect to the chemical 
potential.

Finally, the authors of \cite{kawai} give 
some comments concerning the nature of ``baby universes''
in the theory of branched polymers are which are supposed to
support the claims made. We point out that these speculations are mistaken.

\section{Explanation}
Let us introduce some notation. The random walk representation 
of the Euclidean propagator is given by
\beq\label{4}
G_\m(x_1,x_2) = \sum_{N=1}^\infty e^{-\m N} \int \prod_{i=1}^{N} dy_i\;
\prod_{i=0}^{N}f(y_{i+1}-y_i),~~~y_{N+1}=x_2,~y_0=x_1,
\eeq
where $f(x)$ is a suitable weight function which should fall off sufficiently
fast.
We call \rf{4} the grand canonical partition function for the random walk,
$\m$ the chemical potential and $e^{-\m}$ the fugacity. There is a critical 
value, $\m_c$, for $\m$ above which the sum in \rf{4} is convergent and below which it is divergent. We can write
\beq\label{5}
G_\m(x_1,x_2) = \sum_N e^{-\Del \m N} G_N(x_1,x_2),~~~~\Del \m = \m-\m_c,
\eeq
and $G_N(x_1,x_2)$ is called the canonical partition function for the 
random walk, since the ``internal'' volume (number of steps) 
$N$ is fixed. By Fourier 
transformation we define $G_\m(p)$ and $G_N(p)$. Close to the critical 
point $\m_c$ we have
\beq\label{6}
G_\m(p) \sim \frac{1}{\Del \m +p^2},~~~~G_N(p) \sim e^{-p^2N},
\eeq
such that
\beq\label{7}
G_\m(p) \sim \int dN \; e^{-\Del \m N} G_N(p).
\eeq
It is clear that the relation \rf{1} is not satisfied for random walks, 
and there is no reason
why it should be satisfied, since the lefthand side is the heat kernel and the
righthand side is the propagator.

The value of the Hausdorff dimension of the random walk follows from \rf{7}. 
Without 
going into a detailed derivation (which e.g.\ can be found in \cite{book}),
we simply note that $N$ and $\Del \m$ as well as  $p$ and $x$   
are conjugate variables. In the scaling limit this leads to 
\beq\label{h1}
\la N \ra_\m \sim \frac{1}{\Del \m},~~~|x| \sim \frac{1}{(\Del \m)^{\oh}}
~~~~~~~{\rm i.e.}~~~\la N \ra_\m \sim x^2,
\eeq
where $\la\cdot\ra_\m$ is the expectation in an ensemble of walks whose
endpoints are separated by a distance $x$ in imbedding space.

In the case of branched polymers one obtains similarly 
\cite{book} for $\m$ close to $\m_c$: 
\beq\label{8}
G_\m(p) \sim \frac{1}{\sqrt{\Del \m}+p^2}.
\eeq
This result is universal (as for the random walk) and contains
no non-scaling part. Standard arguments \cite{book}, identical to the ones 
leading to the Eqs.\ \rf{h1} for the random walk, imply (in the scaling limit):
\beq\label{h2}
\la N \ra_\m \sim \frac{1}{\Del \m},~~~
|x| \sim \frac{1}{(\Del \m)^{\oq}},
~~~~~{\rm i.e.}~~~\la N \ra_\m \sim x^4,
\eeq
which shows that the Hausdorff dimension of branched polymers is four.

As for the random walk the 2-point function $G_\m$ for branched polymers 
does not 
satisfy \rf{1}, and for the same reason as before this 
fact does not disqualify it as the correct propagator.
Of course the large-$p$ behaviour of derivatives of $G_\m(p)$ w.r.t. $\m$ is 
\beq\label{9}
\frac{d^l G_\m(p)}{d\m^l} \sim \frac{1}{(\Del \m)^{l-\oh} p^4}+ O(p^{-6}),
\eeq
for $l >0$. The fact that this large-$p$ behavior of \rf{9} agrees with the
large-$p$ behavior of $G_N(p)$ derived from \rf{8} does not make
it the correct propagator for branched polymers. In fact, it is simply
the propagator for branched polymers with $l$ marked vertices since each
differentiation brings down a factor of $N$ which is the number of ways we
can choose a vertex to mark.
  
Next let us comment on Eqs.\ \rf{2} and \rf{3}.
The derivation of \rf{2}, as presented in \cite{book}, is 
acknowledged in \cite{kawai}, but the authors object that 
it does not satisfy \rf{1}. We repeat that it is not 
a relevant objection as we have explained, 
and contrary to their statements there are no
non-universal terms contributing to this equation. 

Eq.\ \rf{3} is an immediate consequence of eq.\rf{2} by differentiating
w.r.t.\ $\m$ on both sides. Indeed, it, and more generally
eq.\ (52) of \cite{kawai}, can also be obtained
before taking the scaling limit by writing the factor $N$ coming from the
differentiation as a sum of the number of vertices associated with each
propagator in the appropriate $\phi^3$-graph, taking care of the
endpoint contributions for the propagators. In particular,  \rf{3} is
by definition not the 3-point function for branched polymers, but 
the 3-point function for branched polymers with one additional marked point.

Let us finally comment on the remarks made in \cite{kawai}
about baby universes. It is claimed that one has a situation 
similar to two-dimensional gravity where a typical surface will 
consist of a ``parent'' universe dressed with small baby 
universes that are connected to the parent by a bottleneck.  
The argument is based on a relation for the 1-point
function\footnote{For some reason Eq.\ (55) in \cite{kawai} misses
a combinatorial factor $N'$, leading to erroneous conclusions.}:
\beq\label{10}
G^1_N > \int dN'\; N' G_{N'} G^1_{N-N'}.
\eeq
However, this relation is only (approximately) 
valid when the entropy exponent 
$\g <0$. In fact, it is usually used to ``derive'' that $\g <0$
in 2d gravity. For branched polymers $\g =1/2$ and \rf{10} is not 
valid (inserting $G^{1}_N \sim N^{\g-2}=N^{-3/2}$ clearly shows 
that \rf{10} is violated for branched polymers (and for all $\g >0$)).
The reason is that {\it if} there are numerous ``baby universes'' of all
sizes, then \rf{10} cannot hold, because the decomposition
of a graph with one marked point into a graph with two marked points of size
$N'$ and a graph with one marked point of size $N-N'$ is not unique. A more
refined treatment \cite{book} leads to the conclusion that  $\g > 0$ implies $\g
=1/2$ under very general conditions. 
The scenario with 
baby universes of all sizes when $\g >0$ has been verified in numerous computer
simulations (using, by the way, the canonical ensemble!).

\section*{Acknowledgement} J.A. and B.D. thank MaPhySTo -- Centre for 
Mathematical Physics and Stochastic, funded by a grant from The Danish 
National Research Foundation -- for support.

\end{document}